\documentclass[12pt]{article}
\usepackage{amsmath,natbib}
 \usepackage[a4paper, total={6in, 8in}]{geometry}

\begin{document}

\title{Bayesian Estimation of Attribute and Identification Disclosure Risks in Synthetic Data}
\author{Jingchen Hu\footnote{Department of Mathematics and Statistics, Vassar College, Poughkeepsie, NY 12604, USA, jihu@vassar.edu}}
\date{}

\maketitle

\begin{abstract}
The synthetic data approach to data confidentiality has been actively researched on, and for the past decade or so, a good number of high quality work on developing innovative synthesizers, creating appropriate utility measures and risk measures, among others, have been published. Comparing to a large volume of work on synthesizers development and utility measures creation, measuring risks has overall received less attention. This paper focuses on the detailed construction of some Bayesian methods proposed for estimating disclosure risks in synthetic data. In the processes of presenting attribute and identification disclosure risks evaluation methods, we highlight key steps, emphasize Bayesian thinking, illustrate with real application examples, and discuss challenges and future research directions. We hope to give the readers a comprehensive view of the Bayesian estimation procedures, enable synthetic data researchers and producers to use these procedures to evaluate disclosure risks, and encourage more researchers to work in this important growing field.

\end{abstract}


\section{Introduction}

Statistical agencies collect microdata of respondents (individuals or business establishments) through various censuses and surveys. The agencies then make some versions of the collected microdata publicly available, subject to privacy and confidentiality protection (e.g. Title 13 and Title 26, U. S. Code). The agencies need to protect the identity of the respondents, as well as the original attribute information of the respondents which are deemed sensitive. These correspond to identification disclosure and attribute disclosure respectively.

To provide such protection, at the very least, unique identifiers such as Social Security Number (SSN) for individuals and Employer Identification Number (EIN) for business establishments cannot be released in the publicly available microdata. Moreover, other seemingly safe attributes cannot be released at the same time either, because a combination of a small number of attributes would increase the chance of respondent identification greatly. \citet{Sweeney2000} demonstrated that using 1990 U. S. Census summary data, 87\% (216 million out of 248 million) of the US population had reported characteristics that likely made them unique based only on \{5-digit ZIP, gender, date of birth\}, and about half (53\%) are likely to be uniquely identified by only \{place, gender, date of birth\}. 

The statistical agencies thus need to mask the microdata before public release. These masking techniques are called Statistical Disclosure Limitation (SDL) techniques, which include i) data swapping, ii) adding random noise, and iii) micro-aggregation, among others. \citet{SDC2012book} provides a comprehensive review of the SDL techniques for microdata. Though these methods or combinations of them could provide some level of privacy protection, the utility of the masked data (e.g. results from a regression analysis using the masked data should be close to that using the original confidential data) are compromised \citep{RaghuReiterRubin2003}. Moreover, for large and complex surveys, such SDL techniques need to be applied at a high intensity, which is time-consuming and damages the utility of final masked data.

One alternative to the SDL techniques is synthetic data. Based on the theory and applications of multiple imputation methodology for missing data problems \citep{Rubin1987book}, multiply-imputed synthetic data can be generated from the statistical models estimated from the original confidential data. Carefully designed statistical models could produce high utility, low risks public microdata. Multiple synthetic datasets should be generated, and appropriate combining rules have been developed to provide accurate point estimates and variance estimates of parameters of interest. Refer to \citet{ReiterRaghu2007, Drechsler2011book} for details of the combining rules.

Synthetic data come in two flavors, i) partially synthetic, where only sensitive attributes of all or some of the records are synthesized \citep{Little1993}, and ii) fully synthetic, where all attributes of every record are synthesized \citep{Rubin1993}. Since their proposals, great amount of research has been done on developing synthesizers, evaluating the utility and risks of the synthetic data. Refer to \citet{Drechsler2011book} for a comprehensive review of partially and fully synthetic data and their applications. 

It is worth noting that the U. S. Census Bureau has been involved in providing access to microdata through synthetic data products. Examples of their synthetic data products include i) OnTheMap (based on \citet{OnTheMap2008}), ii) the synthetic Longitudinal Business Database - SynLBD (based on \citet{SynLBD2011, SynLBD2014}), iii) synthetic Survey of Income and Program Participation - SIPP Synthetic Beta (based on \citet{SIPPdocument}), among others. Germany also has implemented synthetic versions of their German IAB Establishment Panel (based on \citet{DrechslerBenderRassler2008, DrechslerDundlerBenderRasslerZwick2008}). More and more statistical agencies have started experimenting with synthetic data for their microdata releases.

Among the published work on synthetic data, many focus on developing synthesizers and proposing utility measures (both the global utility measures \citep{KarrKohnenOganianReiterSanil2006} for synthetic data in general, and outcome specific utility measures for the particular application). The risk measures, while important, have been paid less attention to overall. This is understandable for at least three reasons: i) for the attribute disclosure risks (i.e. measuring the probability of an intruder correctly inferring the original values of the synthesized attributes of a respondent), which exists both in partially synthetic data and fully synthetic data, the principled evaluation procedure came out only recently and is not a straightforward procedure to implement; ii) for the identification disclosure risks (i.e. measuring the probability of correctly identifying a respondent by matching with available information from elsewhere), which usually only exist in partially synthetic data, the evaluation procedure can be followed in a straightforward manner, encouraging no further development of the measures themselves; and iii) unlike utility measures, which can vary a lot in different applications, risk evaluation procedures largely depend on the type of risk measure. 

In this paper, we want to present easy-to-follow construction of some Bayesian estimation methods for evaluating attribute disclosure risks and identification disclosure risks. These methods use Bayesian thinking in computing probabilities, which are generally natural and easy to understand. Bayesian probabilities are subjective and Bayesian methods are useful for modeling the beliefs of data intruders. Moreover in the estimation process, different assumptions of intruder's knowledge and behavior can be incorporated at various stages. While flexible and intuitive, the actual implementation of the Bayesian estimation process can be complicated and difficult to execute. Therefore, we aim at highlighting key steps in the estimation process, and complementing with real applications with a focus on the risk evaluation aspect of each application. The readers will also see exciting synthetic data projects for various types of data and protection purposes, and the application-specific disclosure risk measures being considered in each application. Discussion of challenges and future directions are throughout the paper as well as at the end as a summary.

The remainder of paper is organized as follows. In Section \ref{SynRisksOverview} we give an overview of risk evaluation for synthetic data and lay out the two types of disclosure risks we consider, namely the attribute disclosure risks  and identification disclosure risks. Section \ref{attributerisk} presents the Bayesian estimation methods of attribute disclosure risks, from notation, to key estimating steps, then selected examples, and finally discussion and comments. Section \ref{identificationrisk} follows a similar structure, where we present the Bayesian estimation methods of identification disclosure risks. Finally in Section \ref{summary}, we give a summary.

\section{Overview of synthetic data risks evaluation}
\label{SynRisksOverview}

This paper considers two types of disclosure risks: i) attribute disclosure risks, and ii) identification disclosure risks. Depending on the synthetic flavor (partially versus fully), one or both of these two types of disclosure risks potentially exist. In this section, we briefly discuss why for partially synthetic data, both attribute disclosure and identification disclosure risks potentially exist, whereas for fully synthetic data, only attribute disclosure risks are considered and evaluated.

We note that these two types of disclosure risks are generic, i.e. for any synthetic data product, one or both types should be considered and evaluated. Attribute disclosure risks in particular, come in various forms depending on the nature of the synthesized attributes and the type of privacy protection. For some applications, such as \citet{HuReiterWang2014PSD} where fully synthetic individual records were generated, the attribute disclosure risks are in the form of correctly inferring the attributes of a record. In other applications, for example when generating synthetic geolocation applications, researchers had created attribute disclosure risk measures based on distance between synthesized geolocations and actual geolocations \citep{WangReiter2012, Paiva2014SM, QuickHolanWikleReiter2015SS, QuickHolanWikle2018JRSSA}. Moreover, for synthetic business establishment data applications, researchers had created attribute disclosure risk measures based on percentages of closest match \citep{DomingoMateoTorra2001, KimKarrReiter2015JOS}, and other measures based on relative difference between the true largest value and the intruder's estimate \citep{KimReiterKarr2018JAS}. We note that not all of these application-specific disclosure risk measures undergo similar estimation procedures as the ones we present in this paper.

\subsection{Disclosure risks of releasing partially synthetic data}
\label{SynRisksOverview:partially}

In partially synthetic data, only sensitive attributes of all or some of the records are synthesized and some other attributes are left un-synthesized. When some of the un-synthesized attributes are available to the intruder via external databases, a matching mechanism based on the common available attributes may allow the intruder to identify records in the released dataset, thus resulting in identification disclosure risks. For example, suppose a partially synthetic dataset contains 1000 individual records and 6 attributes. Among them, 3 are synthesized sensitive attributes \{age, date of birth, annual income\} and 3 are un-synthesized attributes \{gender, marital status, county\}. Suppose now an intruder knows that person $X$ is in the sample, and the intruder also knows the gender and the county of person $X$. Since both gender and county are un-synthesized, the intruder will have a reasonable chance of identifying person $X$ in the sample. 

In addition to potential identification disclosure risks, attribute disclosure risks exist in partially synthetic data. We can easily imagine an intruder trying to infer the true values of the synthesized attributes given the released synthetic data, the un-synthesized attributes and other information. The availability of the un-synthesized attributes may greatly increase the chance of accurate inference of the synthesized attributes. Continuing the example of person $X$ from earlier, with a possible identification, the intruder can now move to find out the original values of the synthesized age, date of birth and annual income of person $X$. 

\subsection{Disclosure risks of releasing fully synthetic data}
\label{SynRisksOverview:fully}

Several authors had claimed that in fully synthetic data, identification disclosure risks are not applicable since there is no unique mapping of the records in the synthetic data to the records in the original data \citep{HuReiterWang2014PSD, WeiReiter2016}. This is generally true because all attributes of all records are synthesized in fully synthetic data. However, it depends on how the fully synthetic data are generated, and we now turn to two different methods to generate fully synthetic data, and their implications on disclosure risks evaluation. 

The first method was proposed by \citet{Rubin1993}. It treats all units that did not participate in the survey as missing data. It first multiply imputes those missing values to generate synthetic populations and second disseminates synthetic samples from these populations \citep{DrechslerPSD2018}. With this method, there is no correspondence between a fully synthetic record and a real world record with subscript $i$, therefore there is no one-to-one correspondence and subsequently no identification disclosure risks. 

The second method was proposed by \citet{Drechsler2011WS}, who pointed out that one can use the partially synthetic approach \citep{Little1993} to replace all values in the data by synthetic values, also obtaining a fully synthetic dataset \citep{DrechslerPSD2018}. While there could be a one-to-one correspondence between the original and synthetic datasets, we do not assume such information known by the intruder, therefore no identification disclosure risks exist. The selected examples in Section \ref{attributerisk:examples} generated fully synthetic data using the second method.

Although identification disclosure risks are treated as non-existing in fully synthetic data, regardless of the generation methods mentioned above, attribute disclosure risks potentially exist, as an intruder can try to use the released synthetic data and any other information to infer an entire record. For example, suppose a fully synthetic dataset contains 1000 individual records and 6 attributes, \{age, date of birth, annual income, gender, marital status, county\}, of which all are synthesized by the second method (i.e. extending partial synthesis to full synthesis, therefore there exists a one-to-one correspondence). Suppose the intruder has information of all attributes of every record in the published synthetic data except for person $X$, and the intruder knows that person $X$ is in the published synthetic data. Then the intruder might be able to find out the attributes of person $X$ by inferring from known information of all other records.

%

\subsection{Disclosure probabilities and their summaries}

The Bayesian estimation methods we present in this paper focus on calculating the probabilities of attribute disclosure and identification disclosure. For example, for attribute disclosure risks, we present how to estimate the probability of correctly inferring the original values of the synthesized vector of attributes $\mathbf{Y}_i^{s}$ of record $i$ to be $\mathbf{y^*}$ given the synthetic data, the un-synthesized attributes, and any other information. 

In addition, synthetic data disseminators can provide file-level summaries of these record-level probabilities. These file-level summaries can be different depending on the applications. For example, when synthesizing fully categorical data as in \citet{HuReiterWang2014PSD, HuReiterWang2018BA} in Section \ref{attributerisk:examples:fullcat}, ranks and re-normalized probabilities are reported as file-level summaries because of the nature of the synthesis (full) and the nature of the attributes (all categorical). When synthesizing partially continuous data as in \citet{WangReiter2012} in Section \ref{attributerisk:examples:parcon}, Euclidean distances between the intruder's inferred value of the longitude and latitude and the actual longitude and latitude, and subsequent counts within a radius are reported as file-level summaries because of the nature of the synthesis (partial) and the nature of the attributes (continuous, in particular, the longitude and latitude of a record). 

For attribute disclosure, we focus on calculating the record-level disclosure probabilities in the key estimating steps (Section \ref{attributerisk:keysteps}). Then in the application section (Sections \ref{attributerisk:examples}), we present file-level disclosure probability summaries for each application, though we touch on the record-level disclosure probabilities calculation assumptions briefly. For identification disclosure, a standard set of 3 file-level summaries is reported in all selected examples. Therefore, we present both the methods to calculate the record-level disclosure probabilities and the methods to calculate the file-level disclosure probability summaries in the key estimating steps (Section \ref{identificationrisk:keysteps}). We want to make the readers aware of the distinction between the record-level probabilities and file-level summaries of disclosure risks.

\section{Bayesian estimation of attribute disclosure risks}
\label{attributerisk}

As discussed previously with the examples in Section \ref{SynRisksOverview:partially} and Section \ref{SynRisksOverview:fully}, attribute disclosure risks potentially exist for fully synthetic data and partially synthetic data. 

Evaluating the attribute disclosure risks in fully synthetic data had been a seemingly impossible task for a while. Empirical matching or misclassification-based approaches \citep{ShlomoSkinner2010AAS} cannot be used since there is no correspondence between the original and the synthetic datasets. \citet{Skinner2012ISRreview} called for further research on the existing Bayesian approaches to disclosure risk assessment, especially to emphasize the Bayesian thinking rather than simply using the Bayesian machinery in the assessment process. As a response to the call of \citet{Skinner2012ISRreview},  \citet{Reiter2012discussion}  tried to propose principled Bayesian estimation procedure for attribute disclosure risks for fully synthetic data. The general framework laid out in \citet{Reiter2012discussion} was extended by and further developed by \citet{Reiter2014framework} for both fully and partially synthetic data. This general framework gives interpretable probability statements of the attribute disclosure risks, and provides flexible incorporation of different assumptions of the intruder's knowledge and behavior. 

In this section, we present the framework of \citet{Reiter2014framework} for Bayesian estimation of attribute disclosure risks. We use similar notations, highlight the key steps, and illustrate with selected examples. We have chosen these examples that are built upon the framework but tailored for specific purposes and needs of the applications. To be as comprehensive as possible, we focus on the following: i) fully synthetic categorical data \citep{HuReiterWang2014PSD, HuReiterWang2018BA}, ii) partially synthetic continuous data \citep{WangReiter2012}, and iii) fully synthetic count data \citep{WeiReiter2016}. In the end, we will discuss the challenges and future directions of this framework.

\subsection{Notations and setup}
\label{attributerisk:notations}

Let $\mathbf{y}_i = (y_{i1}, \cdots, y_{ip})$ be the vector response of observation $i$ in the original confidential dataset, where direct identifiers (such as name or SSN) are removed. When needed, we use $j$ as the variable index, and $j = 1, \cdots, p$. Among the $p$ variables, i) some are synthesized, denoted by $\mathbf{y}_i^{s}$; and ii) others are un-synthesized, denoted by $\mathbf{y}_i^{us}$. We have $\mathbf{y}_i = (\mathbf{y}_i^{s}, \mathbf{y}_i^{us})$ for the $i$-th observation with its original values, and $\mathbf{y} = (\mathbf{y}^{s}, \mathbf{y}^{us})$ for the entire dataset containing $n$ observations with their original values. We note that when full synthesis is carried out, $\mathbf{y}^{us} = \O$, therefore $\mathbf{y} = \mathbf{y}^{s}$. Without loss of generality, we use $\mathbf{y} = (\mathbf{y}^{s}, \mathbf{y}^{us})$ when introducing the notations, setup and key estimating steps.

On the agency side, $m > 1$ synthetic datasets are released, denoted by $\mathbf{Z}$. Each synthetic dataset is denoted by $\mathbf{Z}_{(l)}$ where $l = 1, \cdots, m$. See \citet{Drechsler2011book} for review of synthetic data and the references therein.

On the intruder side, suppose the intruder intends to learn $\mathbf{y}_i^{s}$ for some record $i$ in $\mathbf{y}$. Several pieces of information can be available to the intruder: i) $\mathbf{y}^{us} = \{\mathbf{y}_i^{us}: i = 1, \cdots, n\}$, the un-synthesized values of all $n$ observations; ii) $A$, any auxiliary information known by the intruder about records in $\mathbf{y}$; and iii) $S$, denoting any information known by the intruder about the process of generating $\mathbf{Z}$. We will discuss each piece in detail in Section \ref{attributerisk:keysteps}.

Let $\mathbf{Y}_i^{s}$ be the random variable representing the intruder's uncertain knowledge of $\mathbf{y}_i^{s}$. The intruder seeks the distribution of $p(\mathbf{Y}_i^{s} \mid \mathbf{Z}, \mathbf{y}^{us}, A, S)$. We note that $\mathbf{y}^{us}$ is included in $\mathbf{Z}$, but we leave it in the expression for the purpose of expression factorization in later steps.

If $\mathbf{Y}_i^{s}$ is a vector of categorical variables, then probabilities of accurately inferring the confidential values can be calculated through $p(\mathbf{Y}_i^{s} = \mathbf{y^*} \mid \mathbf{Z}, A, S)$, where $\mathbf{y^*}$ is one plausible combination of categorical responses of those variables. The examples \citep{HuReiterWang2014PSD, HuReiterWang2018BA} in Section \ref{attributerisk:examples:fullcat} on fully synthetic categorical data are illustrations for $\mathbf{Y}_i^{s}$ being a vector of categorial variables. If $\mathbf{Y}_i^{s}$ is one or multiple continuous or count variables, context-specific file-level attribute disclosure probability summaries should be developed to summarize the attribute disclosure risks. The examples of \citet{WangReiter2012} on partially synthetic continuous data in Section \ref{attributerisk:examples:parcon}, and \citet{WeiReiter2016} on fully synthetic count data in Section \ref{attributerisk:examples:fullcon} provide illustrations for these types of $\mathbf{Y}_i^{s}$. 

For the agency, it is paramount to model different intruder's knowledge and behavior, i.e. assumptions on the level of knowledge of $\mathbf{y}^{us}, A$, and $S$. The framework in \citet{Reiter2014framework} allows the incorporation of these different assumptions at multiple stages in the estimating process, thus giving extensive flexibility to parties trying to evaluate attribute disclosure risks.

\subsection{Key estimating steps}
\label{attributerisk:keysteps}

The intruder seeks the distribution of $p(\mathbf{Y}_i^{s} = \mathbf{y^*} \mid \mathbf{Z}, \mathbf{y}^{us}, A, S)$, where $\mathbf{y^*}$ is one possible inferred value of $\mathbf{Y}_i^{s}$ by the intruder. Recall that $\mathbf{y}^{us}$, $A$, and $S$ are available to the intruder. According to Bayes rule,

\begin{eqnarray}
p(\mathbf{Y}_i^{s}  = \mathbf{y^*} \mid \mathbf{Z}, \mathbf{y}^{us}, A, S)
&\propto& p(\mathbf{Z} \mid \mathbf{Y}_i^{s} = \mathbf{y^*} , \mathbf{y}^{us}, A, S) p(\mathbf{Y}_i^{s} = \mathbf{y^*}  \mid \mathbf{y}^{us}, A, S),
\label{eq:AttriBayesRule}
\end{eqnarray}
where $p(\mathbf{Z} \mid \mathbf{Y}_i^{s} = \mathbf{y^*} , \mathbf{y}^{us}, A, S)$ is the synthetic data distribution given what the intruder knows, and $p(\mathbf{Y}_i^{s} = \mathbf{y^*}  \mid \mathbf{y}^{us}, A, S) $ represents the intruder's prior on $\mathbf{Y}_i^{s} = \mathbf{y^*} $ given $\mathbf{y}^{us}$, $A$, and $S$.

The estimation procedure of Equation (\ref{eq:AttriBayesRule}) varies by the variable type of $\mathbf{Y}_i^{s}$, and assumptions on the level of knowledge of $\mathbf{y}^{us}, A$, and $S$, among other things. Here we go through each of these quantities and their implications in the estimating process, highlight several common practices that have been adopted, before we illustrate with a selection of attribute disclosure risk assessment demonstrations with real synthetic data applications in Section \ref{attributerisk:examples}.

\subsubsection{Knowledge of $\mathbf{y}^{us}$}
\label{attributerisk:keysteps:1}

Recall $\mathbf{y}^{us} = \{\mathbf{y}_i^{us}: i = 1, \cdots, n\}$ is the set of un-synthesized values of all $n$ observations. As mentioned before, when $\mathbf{Z}$ is partially synthetic, since intruder has access to $\mathbf{Z}$, $\mathbf{y}^{us}$ can be determined and thus available. When $\mathbf{Z}$ is fully synthetic, $\mathbf{y}^{us} = \O$, therefore we can drop this term and further simplify the expression for fully synthetic $\mathbf{Z}$ as

\begin{eqnarray}
p(\mathbf{Y}_i^{s}  = \mathbf{y^*}  \mid \mathbf{Z}, A, S)
&\propto& p(\mathbf{Z} \mid \mathbf{Y}_i^{s}  = \mathbf{y^*}, A, S) p(\mathbf{Y}_i^{s}  = \mathbf{y^*} \mid A, S).
\label{eq:AttriBayesRuleFull}
\end{eqnarray}

Often times $\mathbf{Y}_i^{s}$ is further simplified to $\mathbf{Y}_i$, as in \citet{HuReiterWang2014PSD, HuReiterWang2018BA}. However, without loss of generality, we keep $\mathbf{y}^{us}$ in the following discussion. 

\subsubsection{Assumptions about $A$}
\label{attributerisk:keysteps:2}

We use $A$ to denote auxiliary information known by the intruder about records in $\mathbf{y}$. As $\mathbf{y}^{us}$ is either known in partial synthetic or dropped in fully synthetic, $A$ specifically refers to information about $\mathbf{y}^{s}$, the synthesized values. When the intruder seeks $p(\mathbf{Y}_i^{s}  = \mathbf{y^*} \mid \mathbf{Z}, \mathbf{y}^{us}, A, S)$, the distribution of record $i$'s synthesized variables, there are numerous possible scenarios of what the intruder knows about the synthesized values of every other record, denoted by the matrix $\mathbf{y}_{-i}^{s}$. First proposed in \citep{Reiter2012discussion}, a ``worst case" scenario where the intruder knows the original values of the synthesized variables of all records except for record $i$ has been a common practice, i.e. $A = \mathbf{y}_{-i}^{s}$. This practice has been recognized as strong intruder knowledge and conservative, as in many contexts, it is impossible for the intruder to know $\mathbf{y}_{-i}^{s}$. However, it has also been argued that if disclosure risks under such conservative assumption are acceptable, disclosure risks should be acceptable for weaker $A$ \citep{Reiter2012discussion, Reiter2014framework}. As \citet{Reiter2012discussion, HuReiterWang2014PSD} noted, assuming the intruder knows all records but one is related to, but quite distinct from, the assumptions used in differential privacy \citep{Dwork2006DP}. \citet{McClureReiter2012DPandSyn} designed simulation studies to compare the two paradigms.

\subsubsection{Assumptions about $S$}
\label{attributerisk:keysteps:3}

$S$ denotes any information known by the intruder about the process of generating $\mathbf{Z}$. Examples include code for the synthesizer and descriptions of the synthesis model. Such information sometimes can be publicly available with great details. For example, the Census Bureau's Survey of Income and Program Participation (SIPP) Synthetic Beta product has an accompanying document \citet{SIPPdocument} describing their synthesizing process. From the document, we gather that they implemented a Sequential Regression Multivariate Imputation (SRMI) framework, with three main models (linear regression, logistic regression, and Bayesian bootstrap \citep{Rubin1981BayesianBootstrap}) for missing data imputation and synthetic data generation. Such publicly available detailed information should be assumed known by the intruder.

\subsubsection{Choosing the prior $p(\mathbf{Y}_i^{s}  = \mathbf{y^*} \mid \mathbf{y}^{us}, A, S)$}
\label{attributerisk:keysteps:4}

Determining the intruder's prior beliefs of $p(\mathbf{Y}_i^{s}  = \mathbf{y^*} \mid \mathbf{y}^{us}, A, S)$ is another nearly impossible task. \citet{Skinner2012ISRreview} challenged the use of prior distributions being a more technical one for the Bayesian machinery to function, and advocated for prior distributions that should be defensible from the agency's perspective. A common practice is to specify uniform prior distributions of $\mathbf{Y}_i^{s}  = \mathbf{y^*}$ for all possible inferred values $\mathbf{y^*}$, given $(\mathbf{y}^{us}, A, S)$ as proposed in \citet{Reiter2012discussion}, but also consider a variety of prior distributions when possible, especially if more informative prior is available \citep{WeiReiter2016}.

\subsubsection{The estimation of $p(\mathbf{Z} \mid \mathbf{Y}_i^{s} = \mathbf{y^*}, \mathbf{y}^{us}, A, S)$}
\label{attributerisk:keysteps:5}

We now go through the estimation of $p(\mathbf{Z} \mid \mathbf{Y}_i^{s} = \mathbf{y^*}, \mathbf{y}^{us}, A, S)$. The previously discussed assumptions of $A$ and $S$ are relevant in this part of the estimating process. The importance sampling techniques coupled with Monte Carlo simulation are adopted common practices, and we will present and discuss why and how they work.

Typically by the independence of $m$ different synthetic datasets, we work with each synthetic dataset $\mathbf{Z}_{(l)}$ separately, therefore we consider $p(\mathbf{Z}_{(l)} \mid \mathbf{Y}_i^{s} = \mathbf{y^*}, \mathbf{y}^{us}, A, S)$ for our discussion. To ultimately obtain $p(\mathbf{Z} \mid \mathbf{Y}_i^{s} = \mathbf{y^*}, \mathbf{y}^{us}, A, S)$, we have

\begin{eqnarray}
p(\mathbf{Z} \mid \mathbf{Y}_i^{s} = \mathbf{y^*}, \mathbf{y}^{us}, A, S) = \prod_{l=1}^{m} p(\mathbf{Z}_{(l)} \mid \mathbf{Y}_i^{s} = \mathbf{y^*}, \mathbf{y}^{us}, A, S).
\label{eq:AttriProd}
\end{eqnarray}

For $p(\mathbf{Z}_{(l)} \mid \mathbf{Y}_i^{s} = \mathbf{y^*}, \mathbf{y}^{us}, A, S)$, under the ``worst case" scenario where the intruder knows the original values of the synthesized variables of all records except for record $i$, i.e. $A = \mathbf{y}_{-i}^{s}$, we come to 

\begin{eqnarray}
p(\mathbf{Z}_{(l)} \mid \mathbf{Y}_i^{s}  = \mathbf{y^*}, \mathbf{y}^{us}, A = \mathbf{y}_{-i}^{s}, S),
\label{eq:AttriImportance1}
\end{eqnarray}
which is very close to the distribution from which the synthetic data $\mathbf{Z}_{(l)}$ is generated, as in

\begin{eqnarray}
p(\mathbf{Z}_{(l)} \mid \mathbf{y}, S) = p(\mathbf{Z}_{(l)} \mid \mathbf{Y}_i^{s} = \mathbf{y}_i, \mathbf{y}^{us}, A = \mathbf{y}_{-i}^{s}, S), 
\label{eq:AttriImportance2}
\end{eqnarray}
where $\mathbf{y}_i$ is the true record in the original confidential dataset $\mathbf{y}$. As we can see, the only difference in the conditioned quantities in Equation (\ref{eq:AttriImportance1}) and Equation (\ref{eq:AttriImportance2}) is the difference between $\mathbf{y^*}$ (the random inferred value) and $\mathbf{y}_i$ (the true record). 

In fact, we could utilize the small difference between $\mathbf{y}$ and $(\mathbf{Y}_i^{s} = \mathbf{y^*}, \mathbf{y}^{us}, A = \mathbf{y}_{-i}^{s}, S)$ for estimating $p(\mathbf{Z}_{(l)} \mid \mathbf{Y}_i^{s} = \mathbf{y}^*, \mathbf{y}^{us}, A = \mathbf{y}_{-i}^{s}, S)$. If we use $\Theta$ to denote the parameters in the synthesis model $M$, we could easily incorporate $\Theta$ draws in our estimation of $p(\mathbf{Z}_{(l)} \mid \mathbf{Y}_i^{s} = \mathbf{y}^*, \mathbf{y}^{us}, A = \mathbf{y}_{-i}^{s}, S)$ through a Monte Carlo step, as in

\begin{eqnarray}
p(\mathbf{Z}_{(l)} \mid \mathbf{Y}_i^{s} = \mathbf{y}^*, \mathbf{y}^{us}, A = \mathbf{y}_{-i}^{s}, S) = \int p(\mathbf{Z}_{(l)} \mid \mathbf{Y}_i^{s} = \mathbf{y}^*, \mathbf{y}^{us}, A = \mathbf{y}_{-i}^{s}, S, \Theta) \nonumber \\
p(\Theta \mid \mathbf{Y}_i^{s} = \mathbf{y}^*, \mathbf{y}^{us}, A = \mathbf{y}_{-i}^{s}, S)d \Theta.
\label{eq:AttriImportance3}
\end{eqnarray}

The Monte Carlo step requires re-estimation of the synthesis model $M$ for each $\mathbf{Y}_i^{s} = \mathbf{y^*}$, which could be computationally prohibitive if many possible inferred values of $\mathbf{Y}_i^{s}$ need to be evaluated. To avoid the re-estimation of $M$ to draw $\Theta$ samples, a common procedure via importance sampling is adopted. In particular, available draws of $\Theta$ from $p(\Theta \mid \mathbf{y})$, the model used for generating the synthetic dataset $\mathbf{Z}_{(l)}$, act as proposals for the importance sampling algorithm. Readers are referred to \citet{Paiva2014SM, HuReiterWang2014PSD} for a review of importance sampling and its usage in the applications therein.

\subsection{Selected examples}
\label{attributerisk:examples}
In this selected examples section, we want to show the readers in a few different applications, i) what are $\mathbf{y}^{s}$, $\mathbf{y}^{us}$, $\mathbf{Y}_i^s$, $A$, and $S$; ii) what are the risk scenarios (i.e. assumptions are made for $A$ and $S$), and their implications; and iii) what are the specific file-level attribute disclosure probability summaries in each application. For each application, we give a brief overview of the dataset(s) and research questions to provide the background. We also mention the synthesizers, but the details of the synthesizers and the evaluation of the utility of the synthetic data are omitted, as we focus on the estimation of probabilities of attribute disclosure in this paper. Interested readers should refer to the cited papers for further information.

\subsubsection{Fully synthetic categorical data}
\label{attributerisk:examples:fullcat}
\citet{HuReiterWang2014PSD} aimed at generating fully synthetic categorical data for a subset of $n=10000$ individuals from the 2012 American Community Survey (ACS) public use microdata sample for the state of North Carolina. They considered $14$ unordered categorical variables, as listed in Table \ref{tab:HuReiterWang2014}. We include the variables, the number of categories of each variable, and whether a variable is synthesized in this table.

\begin{table}[bth]
\centering
\caption{Variables used in the \citet{HuReiterWang2014PSD}. Data taken
  from the 2012 ACS public use microdata
  samples. \label{tab:HuReiterWang2014}}
\begin{tabular}{lll}
\hline
Variable &  Number of categories & Synthesized \\ \hline
Sex &  2 & Yes\\
Age & 4 & Yes\\
Race & 6 & Yes \\
Education level & 4 & Yes\\
Marital status &5 & Yes\\
Language & 2 & Yes\\
Birth place & 7 & Yes \\
Military & 3 & Yes \\
Work &  3 & Yes\\
Disability &  2 & Yes\\
Health insurance coverage & 2 & Yes\\
Migration & 3 & Yes\\
School & 3 & Yes\\
Hispanic &2 & Yes\\ \hline
\end{tabular}
\end{table}

While the authors attempted fully synthetic data generation \citep{Rubin1993}, they followed the partially synthetic approach \citep{Little1993} to replace all values in the data by synthetic values \citep{Drechsler2011WS}. Unlike the approach of \citet{Rubin1993}, where no correspondence between a fully synthetic record and a real world record with subscript $i$ exists, their approach maintains such correspondence, which is a key assumption in their proposed attribute disclosure evaluation methods. For more discussion on the two approaches to generating fully synthetic data, refer to \citet{DrechslerPSD2018}.

The authors used a Dirichlet Process mixture of products of multinomial (DPMPM) synthesizer. The DPMPM is consisted of a set of flexible Bayesian latent class models that have been developed to capture complex relationships among multivariate unordered categorical variables \citep{DunsonXing2009JASA}. In recent years, the DPMPM has been proposed as a multiple imputation engine for missing data problems, and a synthesizer for SDL. \citet{SiReiter2013JEBS} implemented the DPMPM as a missing data imputation engine and demonstrated its superior performance comparing to traditional sequential imputation models such as the multiple imputation with chained equations (MICE; \citet{miceR}). In addition to \citet{HuReiterWang2014PSD}, \citet{DrechslerHu2018, HuSavitsky2018} used the DPMPM synthesizer for generating partially synthetic data with geocoding information. Variations of the DPMPM include versions of it dealing with structural zeros \citep{ManriqueReiter2014JCGS, ManriqueHu2018JRSSA}, dealing with extension of the multinomial synthesizer \citep{HuHoshinoPSD2018}, and dealing with individuals nested within households \citep{HuReiterWang2018BA, AkandeReiterBarrientos2018SM, AkandeBarrientosReiter2018JSSAM}.

The DPMPM synthesizer assigns an underlying latent class of each record. Conditioning on the latent class assignment, each attribute independently follows its own distribution. For unordered categorical variables, such distribution is usually multinomial distribution. To generate the synthetic vector of attributes of one record, we first sample the latent class assignment. For each attribute, we generate the value from its independent multinomial distribution with probabilities sampled from the DPMPM. 

We use $\mathbf{Y}_i$ to represent the random attribute vector of record $i$ (the superscript $s$ is dropped because this is fully synthetic, i.e. $\mathbf{y}_i^{us} = \O$ and $\mathbf{Y}_i^{s} = \mathbf{Y}_i$), $\mathbf{Y}$ to represent the random attribute vectors of all $n$ records, $\mathbf{Z}_{(l)}$ to represent each fully synthetic dataset, where $l = 1, \cdots, m$, and $\mathbf{Z} = \{\mathbf{Z}_{(1)}, \cdots, \mathbf{Z}_{(m)}\}$ to represent all $m$ fully synthetic datasets.

Following the general setup and notations introduced earlier, we use $A$ to represent the intruder's information on person's attributes in the sample (i.e. auxiliary information), and $S$ to represent any meta-data released by the agency about the synthesis model. The goal is to estimate $\mathbf{Y}_i^{s}$ for one or more target records in the sample. Specifically in this case, because all attributes are unordered categorical, we are able to enumerate all possible combinations of the categorical attributes. 

Then the expression of the probability of attribute disclosure of an entire record $i$ becomes Equation (\ref{eq:AttriBayesRule-rep1}),

\begin{eqnarray}
p(\mathbf{Y}_i = \mathbf{y^*} \mid \mathbf{Z},A, S)
&\propto& p(\mathbf{Z} \mid \mathbf{Y}_i = \mathbf{y^*}, A, S) p(\mathbf{Y}_i = \mathbf{y^*} \mid A, S),
\label{eq:AttriBayesRule-rep1}
\end{eqnarray}
where $\mathbf{y^*}$ is an inferred value by the intruder.

\citet{HuReiterWang2014PSD} set $A = \mathbf{y}_{-i}$, which corresponds to the ``worst case" scenario where the intruder knows the actual data for all records except for the record $i$. 

\begin{eqnarray}
p(\mathbf{Y}_i = \mathbf{y^*} \mid \mathbf{Z},A  = \mathbf{y}_{-i}, S)
&\propto& p(\mathbf{Z} \mid \mathbf{Y}_i = \mathbf{y^*}, A  = \mathbf{y}_{-i}, S) p(\mathbf{Y}_i = \mathbf{y^*} \mid A  = \mathbf{y}_{-i}, S). \nonumber \\
\label{eq:AttriBayesRule-rep2}
\end{eqnarray}

To estimate $p(\mathbf{Z} \mid \mathbf{Y}_i = \mathbf{y^*}, A  = \mathbf{y}_{-i}, S)$, the authors proposed to dramatically reduce the set of possible combinations that $\mathbf{y^*}$ could take. Specifically, they consider the neighborhood near $\mathbf{y}_i$ (the true record), which contains only feasible candidates where $\mathbf{y^*}$ differ from $\mathbf{y}_i$ in one variable.  In their illustrative application, the subset contains only $35$ combinations, reduced from $8709120$ cells in the contingency table. The authors commented that if risks of $\mathbf{Y}_i$ being the true $\mathbf{y}_i$ are acceptable in this reduced set, then the risks would be even lower when considering the full set. The risks in the reduced set are the upper bound. Importance sampling techniques were applied to avoid re-evaluation of probability of obtaining the synthetic datasets $\mathbf{Z}$ given different combinations of $(\mathbf{Y}_i = \mathbf{y^*}, A  = \mathbf{y}_{-i}, S)$, as in Equation (\ref{eq:AttriImportance1}). 

For the prior on $p(\mathbf{Y}_i = \mathbf{y^*} \mid A  = \mathbf{y}_{-i}, S)$, \citet{HuReiterWang2014PSD} assumed a uniform prior, which sums to $1$ over the number of combinations in the reduced set (e.g. $1/35$ in their illustration). The prior probabilities were canceled out in the computation process.

To summarize the calculated risks of attribute disclosure of all records, \citet{HuReiterWang2014PSD} created two file-level attribute disclosure probability summaries: 

\begin{enumerate}
\item[(i)] The ranking of the probability of the true record being disclosed, among the subset of combinations; 
\item[(ii)] The re-normalized probability of the true record being disclosed, among the subset of combinations. 
\end{enumerate}
In general, the higher the ranking and the re-normalized probability, the higher the attribute disclosure risks.

Additionally, \citet{HuReiterWang2014PSD} investigated scenarios where the intruder might know a subset of values in $\mathbf{y}_i$, for example, demographic variables. The authors then defined each $\mathbf{y}_i$ as $\mathbf{y}_i = (\mathbf{y}_{i, k}, \mathbf{y}_{i, uk})$, where the additional subscript $k$ denotes the variables known by the intruder and $uk$ denotes the variables unknown by the intruder. Subsequently, to evaluate risks for intruders seeking to estimate the distribution of $\mathbf{Y}_{i, uk}$, the authors defined $A = (\mathbf{y}_i, \mathbf{y}_{i, k})$, and Equation (\ref{eq:AttriBayesRule-rep2}) becomes

\begin{eqnarray}
p(\mathbf{Y}_{i, uk} = \mathbf{y^*}_{uk} \mid \mathbf{Z}, A = (\mathbf{y}_i, \mathbf{y}_{i, k}), S)
&\propto& p(\mathbf{Z} \mid \mathbf{Y}_{i, uk} = \mathbf{y^*}_{uk} , A = (\mathbf{y}_i, \mathbf{y}_{i, k}), S) \nonumber \\
&&p(\mathbf{Y}_{i, uk} = \mathbf{y^*}_{uk} \mid A = (\mathbf{y}_i, \mathbf{y}_{i, k}), S). 
\label{eq:AttriBayesRule-rep3}
\end{eqnarray}

The estimation procedure works in a similar way, and we refer interested readers to \citet{HuReiterWang2014PSD} for detail and discussion.

\subsubsection{Partially synthetic continuous data}
\label{attributerisk:examples:parcon}

\citet{WangReiter2012} aimed at generating partially synthetic data for sharing precise geographies. The precise geographies were exact longitude and latitude of each death of a sample of $n = 2670$ North Carolina mortality records in 2002. Only exact longitude and latitude of each record were synthesized, and all non-geographic variables were kept unchanged. We include the variables, their descriptions, and whether a variable is synthesized in Table (\ref{tab:WangReiter2012}).

\begin{table}[bth]
\centering
\caption{Variables used in the \citet{WangReiter2012}. Data taken
  from the 2012 North Carolina mortality records dataset. \label{tab:WangReiter2012}}
\begin{tabular}{lll}
\hline
Variable &  Description & Synthesized \\ \hline
Longitude &  Recoded (1 - 100) & Yes\\
Latitude & Recoded (1 - 100) & Yes\\
Sex & Male, female & No \\
Race & White, black & No\\
Age (years) & 16-99 & No\\
Autopsy performed & Yes, no, missing & No\\
Autopsy findings & Yes, no, missing & No \\
Marital status & 5 categories & No \\
Attendant &  3 categories & No\\
Hispanic & 7 categories & No \\
Education (years) & 0-17 & No\\
Hospital type & 8 categories & No\\
Cause of death & Binary & No\\ \hline
\end{tabular}
\end{table}

The authors used classification and regression trees (CART; refer to \citet{Reiter2005JOS} for details of using CART to generate partially synthetic data) synthesizers for generating longitudes and latitudes. In particular, they first fit a regression tree of longitudes on all non-geographic attributes, and generated synthetic longitudes using the Bayesian bootstrap. After obtaining synthetic longitudes, they fit another regression tree of latitude on all non-geographic attributes and the true latitude, and generated synthetic latitudes using the Bayesian bootstrap. In the end, the partially synthetic precise geographies were simulated. 

We use $\mathbf{Y}_i^{s}$ to represent the longitude and latitude of record $i$, $\mathbf{Y}^{s}$ to represent the random longitudes and latitudes of all $n$ records, $\mathbf{y}^{us}$ to represent the un-synthesized value of all $n$ records, $\mathbf{Z}_{(l)}$ to represent each partially synthetic dataset, where $l = 1, \cdots, m$, and $\mathbf{Z} = \{\mathbf{Z}_{(1)}, \cdots, \mathbf{Z}_{(m)}\}$ to represent all $m$ partially synthetic datasets. 

Following the general setup and notations introduced earlier, we use $A$ to represent the intruder's information on person's attributes in the sample (i.e. auxiliary information), and $S$ to represent any meta-data released by the agency about the synthesis model. The goal is to estimate $\mathbf{Y}_i^{s}$ for one or more target records in the sample. We now express the probability of attribute disclosure of estimating the longitude and latitude of record $i$, which is the same as in Equation (\ref{eq:AttriBayesRule}) and restated below in Equation (\ref{eq:AttriBayesRule-rep3}). Recall that $\mathbf{y^*}$ is a possible original value of $\mathbf{Y}_i^{s}$ by the intruder.

\begin{eqnarray}
p(\mathbf{Y}_i^{s} = \mathbf{y^*} \mid \mathbf{Z}, \mathbf{y}^{us}, A, S)
&\propto& p(\mathbf{Z} \mid \mathbf{Y}_i^{s}  = \mathbf{y^*}, \mathbf{y}^{us}, A, S) p(\mathbf{Y}_i^{s}  = \mathbf{y^*} \mid \mathbf{y}^{us}, A, S) 
\label{eq:AttriBayesRule-rep3}
\end{eqnarray}

\citet{WangReiter2012} evaluated two scenarios regarding the choice of $A$ and $S$.

\begin{enumerate}
\item[(i)] Scenario \#1 (high-risk): The intruder knows everything except for one target's $\mathbf{Y}_i^{s}$, i.e., $A = \mathbf{y}_{-i}^{s}$, and $S$ includes everything about the CART except the individual geographies in the nodes. Note that $A = \mathbf{y}_{-i}^{s}$ is the same as the ``worst case" scenario discussed in Section \ref{attributerisk:keysteps}.

\begin{eqnarray}
p(\mathbf{Y}_i^{s}   = \mathbf{y^*} \mid \mathbf{Z}, \mathbf{y}^{us}, A =  \mathbf{y}_{-i}^{s}, S)
&\propto& p(\mathbf{Z} \mid \mathbf{Y}_i^{s}  = \mathbf{y^*}, \mathbf{y}^{us}, A =  \mathbf{y}_{-i}^{s}, S) \nonumber \\
&&p(\mathbf{Y}_i^{s}  = \mathbf{y^*} \mid \mathbf{y}^{us}, A =  \mathbf{y}_{-i}^{s}, S)  \nonumber \\
\label{eq:AttriBayesRule-rep4}
\end{eqnarray}

\item[(ii)] Scenario \#2 (low-risk): The intruder does not know any records' geographies, i.e., $A = \O$, and $S$ includes everything about the CART except the individual geographies in the nodes.

\begin{eqnarray}
p(\mathbf{Y}_i^{s}   = \mathbf{y^*} \mid \mathbf{Z}, \mathbf{y}^{us}, A =  \O, S)
&\propto& p(\mathbf{Z} \mid \mathbf{Y}_i^{s}  = \mathbf{y^*}, \mathbf{y}^{us}, A = \O, S) \nonumber \\
&& p(\mathbf{Y}_i^{s}  = \mathbf{y^*} \mid \mathbf{y}^{us}, A =  \O, S)  \nonumber \\
\label{eq:AttriBayesRule-rep5}
\end{eqnarray}

\end{enumerate}

For the high-risk scenario, to estimate $p(\mathbf{Z} \mid \mathbf{Y}_i^{s} = \mathbf{y^*}, \mathbf{y}^{us}, A = \mathbf{y}_{-i}^{s}, S)$, importance sampling techniques were applied to avoid re-evaluation of probability of obtaining the synthetic datasets $\mathbf{Z}$ given different combinations of $(\mathbf{Y}_i^{s} = \mathbf{y^*}, \mathbf{y}^{us}, A, S)$, as in Equation (\ref{eq:AttriImportance1}). For the intruder's prior on $p(\mathbf{Y}_i^{s} = \mathbf{y^*} \mid \mathbf{y}^{us}, A = \mathbf{y}_{-i}^{s}, S)$, \citet{WangReiter2012} assumed a uniform distribution on a grid over a small area containing the target's true longitude and latitude. 

The authors noted that the uniform prior for $p(\mathbf{Y}_i^{s} = \mathbf{y^*} \mid \mathbf{y}^{us}, A = \mathbf{y}_{-i}^{s}, S)$ represents strong intruder prior information, because the prior was given on a grid over a small area. They also noted that the value of the risk measure would change if other prior specifications were given, though they did not consider other specifications in their attribute risk disclosure evaluation.

Specifically, \citet{WangReiter2012} developed two geographies-specific attribute disclosure risk measures:

\begin{enumerate}
\item[(i)] A Euclidean distance $R_1$ between the intruder's inferred value of the longitude and latitude and the actual longitude and latitude; 
\item[(ii)] The count $R_2$ recording the number of actual cases in circle centered at the actual longitude and latitude with radius $R_1$. 
\end{enumerate}

In general, larger values of $R_1$ and $R_2$ correspond to smaller attribute disclosure risks. 

We also want to note that \citet{Paiva2014SM} used a similar dataset with the goal of partially synthesizing geographies. Though their synthesis model was different from the one in \citet{WangReiter2012}, they proposed three other file-level attribute disclosure probability summaries for synthetic data applications involving geographic locations.

\begin{enumerate}
\item[(i)] A file-level risk measure of the percentage of records with the true location $\mathbf{y}_i$ being the maximum posterior probability of record $i$;
\item[(ii)] A file-level risk measure of the percentage of records with the true location $\mathbf{y}_i$ being the maximum posterior probability of record $i$, and record $i$ has unique patterns;
\item[(iii)] A Euclidean distance measure between the true location $\mathbf{y}_i$ and the inferred value $\mathbf{y^*}$ with the maximum posterior probability of record $i$.
\end{enumerate}

In general, smaller values of measures (i) and (ii) and larger values of measure (iii) correspond to smaller attribute disclosure risks. 

\subsubsection{Fully synthetic continuous data}
\label{attributerisk:examples:fullcon}

\citet{WeiReiter2016} aimed at generating fully synthetic data for sharing magnitude microdata from business establishments. The magnitude variables were the number of skilled laborers, the number of unskilled laborers, wages of skilled laborers, and wages of unskilled laborers of a sample of $n = 1051$ food manufacturing establishments in the country of Colombia in 1977. All 4 magnitude variables were synthesized, making it a fully synthetic microdata endeavor. We include the variables, their descriptions, and whether a variable is synthesized in Table (\ref{tab:WeiReiter2016}).

\begin{table}[bth]
\centering
\caption{Variables used in the \citet{WeiReiter2016}. Data taken
  from the 1977 Colombia food manufacturing establishments sample. \label{tab:WeiReiter2016}}
\begin{tabular}{lll}
\hline
Variable &  Description & Synthesized \\ \hline
Number of skilled laborers &  Integer & Yes\\
Number of unskilled laborers & Integer & Yes\\
Wages of skilled laborers & Integer & Yes\\
Wages of unskilled laborers &  Integer & Yes\\ \hline
\end{tabular}
\end{table}

The authors used three synthesizers based on finite mixtures of Poisson (MP) distributions. The class of finite mixtures of Poissons can i) capture complex multivariate associations among the variables; and ii) model count variables. In addition to the basic MP synthesizer, \citet{WeiReiter2016} proposed the mixture of Multinomial (MM) synthesizer, which ensures the synthetic values sum to marginal totals in the confidential data. The marginal totals constraints are satisfied by performing another layer of Multinomial draws of counts within each occupied Poisson mixture component. Specifically, the totals (e.g. the number of skilled laborers) and the number of cases (in running the MP, at each MCMC iteration, each record is assigned to a component) in each occupied Poisson mixture component are computed and stored. Based on the totals and the number of cases, a Multinomial sample is generated, distributing the totals into all levels. Within each occupied Poisson mixture, the marginal totals match in the synthetic and confidential data, therefore overall marginal totals also match. Furthermore, the authors proposed the tail-collapsed mixture of Multinomial (TCMM) synthesizer, which effectively performs a model-based variation of microaggregation plus noise, by collapsing tails of individual variables (i.e. risky values). For TCMM, one needs to specify a parameter associated with the quantile, namely $q$, which acts as a threshold to control the amount of collapsing.

We use $\mathbf{Y}_i$ to represent the random vector of the numbers of skilled and unskilled laborers and their corresponding wages of record $i$, $\mathbf{Y}$ to represent the 4 random magnitude variables of all $n$ records, $\mathbf{Z}_{(l)}$ to represent each fully synthetic dataset, where $l = 1, \cdots, m$, and $\mathbf{Z} = \{\mathbf{Z}_{(1)}, \cdots, \mathbf{Z}_{(m)}\}$ to represent all $m$ fully synthetic datasets. Note that we dropped the superscript $s$ in $\mathbf{Y}^{s}$ and use $\mathbf{Y}$ and $\mathbf{Y}_i$ directly because every variable is synthesized.

Specific to the business establishment survey data, we need to define a few other quantities before we can describe the risk scenarios and file-level attribute disclosure probability summaries. We use $y_{(1)j}$ and $y_{(2)j}$ to represent the largest and second largest values of variable $j$ in $\mathbf{y}$, the original confidential dataset. When the intruder does not know these values, we use $Y_{(1)j}$ and $Y_{(2)j}$ as the random variables representing the intruder's uncertain knowledge about them. Furthermore, we let $T_j$ be the total of variable $j$ in $\mathbf{y}$, and use $\mathbf{y}_{(1)}$ and $\mathbf{y}_{(2)}$ to represent the values of two entire records.

\citet{WeiReiter2016} considered a variety of risk scenarios. As an illustration of evaluating attribute disclosure for fully synthetic continuous data, we present the scenario where the intruder, who has the second largest value of a certain variable, $y_{(2)j}$, attempts to use the released synthetic data $\mathbf{Z}$ to learn about the individual with the largest value of the variable, the random quantity $Y_{(1)j}$. Such a scenario is commonly used by official statistics agencies with business establishment data \citep{KimKarrReiter2015JOS, KimReiterKarr2018JAS}. 

Recall that we use $A$ to represent the intruder's information on person's attributes in the sample (i.e. auxiliary information), and $S$ to represent any meta-data released by the agency about the synthesis model. Translating the scenario above into choices of $A$ and $S$, we come to Equation (\ref{eq:AttriBayesRule-rep6}), which represents the attribute disclosure risk probability of correctly inferring $Y_{(1)j} = y^*_{(1)j}$ when $T_j$ is available.

\begin{eqnarray}
Pr(Y_{(1)j} = y^*_{(1)j} \mid \mathbf{Z}, \mathbf{y}_{(2)}, Y_{(1)j}  \geq y_{(2)j}, T_j) &\propto& Pr(\mathbf{Z} \mid Y_{(1)j} = y^*_{(1)j}, \mathbf{y}_{(2)}, T_j)  \nonumber \\
&&Pr(Y_{(1)j} = y^*_{(1)j} \mid \mathbf{y}_{(2)}, Y_{(1)j} \geq y_{(2)j}, T_j), \nonumber \\
\label{eq:AttriBayesRule-rep6}
\end{eqnarray}
where $y^*_{(1)j}$ is a possible original value of $Y_{(1)j}$ by the intruder.

To estimate $Pr(\mathbf{Z} \mid Y_{(1)j} = y^*_{(1)j}, \mathbf{y}_{(2)}, T_j)$, techniques of using $\mathbf{Z}$ to approximate the set of records in the same component occupied by the target record were applied to simplify the computation. We refer the readers to \citet{WeiReiter2016} for the details regarding the MM and TCMM synthesizers. 

For the intruder's prior on $Pr(Y_{(1)j} = y^*_{(1)j} \mid \mathbf{y}_{(2)}, Y_{(1)j} \geq y_{(2)j}, T_j)$, \citet{WeiReiter2016} discussed the choice of a non-uniform prior distribution, which could provide more accurate prior inferred values and is worth noting here. In their empirical illustration of synthesizing fully magnitude data from the Colombia food manufacturing establishments dataset, the authors estimated the chance that the largest value of the number of skilled laborers falls into an interval, with the lower bound being the second largest value and the upper bound being pre-defined. Among the three different synthesizers, the attribute disclosure risks under the MP and the MM synthesizers are extremely high, whereas the risks under the TCMM synthesizer are overall much lower. Furthermore, the risks decrease as the threshold parameter $q$ decreases, which is expected because of the amount of tail collapsing increases as $q$ decreases. Interested readers are encouraged to consult \citet{WeiReiter2016} for their explanations.

We also want to point out that \citet{WeiReiter2016} evaluated two other sets of scenarios, both of which assume the intruder seeks to correctly infer the values of variable $j$ of two records, $y_{(1)j}$ and $y_{(2)j}$. The first scenario is the intruder knows all but one or two values in $\mathbf{y}$, which means the intruder seeks to estimate the probability of $(Y_{(1)j} = y^*_{(1)j}, Y_{(2)j} = y^*_{(2)j})$ given $\mathbf{Z}$, $Y_{(1)j} + Y_{(2)j} = T_{2j}$ and different combinations of $A$ and $S$. The second scenario is the intruder knows all data values except for one or two records, which means the intruder seeks to estimate the probability of $(Y_{(1)j} = y^*_{(1)j}, Y_{(2)j} = y^*_{(2)j})$ given $\mathbf{Z}$, $\mathbf{Y}_{(1)} + \mathbf{Y}_{(2)} = \mathbf{T}_{2}$ and different combinations of $A$ and $S$.

\subsection{Discussion and comments}
\label{iattributerisk:discussion}

There are few common practices for evaluating attribute disclosure risks in the selected examples, as well as in other synthetic applications. The first one is on the assumptions about $A$, the auxiliary information known by the intruder about records in $\mathbf{y}^s$. The ``worst case" scenario of letting $A = \mathbf{y}_{-i}^s$, i.e. the intruder knows all the original values of the synthesized variables of all records except for record $i$, although it provides an upper bound on the identification risks, is a very strong and probably unrealistic assumption. The scenario greatly simplifies the estimation of $p(\mathbf{Z} \mid \mathbf{Y}_i^s = \mathbf{y^*}, \mathbf{y}^{us}, A, S)$ as in Section \ref{attributerisk:keysteps:5} and Equation (\ref{eq:AttriProd}), by setting $A = \mathbf{y}_{-i}^s$. If the assumption is weaker, for example, the intruder only knows the synthesized values of next record $i+1$, then $A = \mathbf{y}_{i+1}^s$, which means the approximation in Equation (\ref{eq:AttriImportance1}) to (\ref{eq:AttriImportance3}) will involve extra steps of imputing all the other synthesized values $\{\mathbf{y}_{i'}^s, i' \neq i, i' \neq i+1\}$ (see \citet{Paiva2014SM} for a potential solution). Such weaker assumptions are much more realistic, but almost computationally infeasible with the current setup. \citet{McclureReiter2016} examined the effect on attribute disclosure risks in fully synthetic data by decreasing the number of observations knows (i.e. weakening the assumption of $A = \mathbf{y}_{-i}^s$). Future research in designing faster algorithms to estimate $p(\mathbf{Z} \mid \mathbf{Y}_i^s = \mathbf{y^*}, \mathbf{y}^{us}, A, S)$ with weaker $A$ is desired.

The common practice of setting the prior $p(\mathbf{Y}_i^s = \mathbf{y^*} \mid \mathbf{y}^{us}, A, S)$ as a uniform distribution has been adopted in various applications. Because of the cancellation in Bayes' rule, using a uniform prior for $p(\mathbf{Y}_i^s = \mathbf{y^*} \mid \mathbf{y}^{us}, A, S)$ essentially simplifies the estimation, as we only need to estimate $p(\mathbf{Z} \mid \mathbf{Y}_i^s = \mathbf{y^*}, A, S)$ in Equation (\ref{eq:AttriBayesRuleFull}). We should recognize not only its convenience in computation, but also its constraint. Using a uniform prior can be un-informative for some cases, but it might be strongly informative in other cases, as in \citet{WangReiter2012}, which might not be realistic. Using a uniform prior can be realistic in some cases, but it might need to adjusted to reflect more realistic prior belief. For example, it is possible to argue that the 35 combinations in the reduced subset in \citet{HuReiterWang2014PSD} should not really be treated equally likely (i.e. a uniform prior). Rather, some combinations might be more plausible than the others, thus carrying higher prior probability. The general advice is to consider a wide range of prior distributions for $p(\mathbf{Y}_i^s = \mathbf{y^*} \mid \mathbf{y}^{us}, A, S)$ if possible. Also, do not choose uniform only for its simplicity. Choosing a more realistic prior distribution provides a more reasonable attribute disclosure risks measure \citep{WeiReiter2016}.

When estimating $p(\mathbf{Z} \mid \mathbf{Y}_i^s = \mathbf{y^*}, \mathbf{y}^{us}, A, S)$, importance sampling techniques are widely used to avoid re-estimating the synthesis model $M$ for each $\mathbf{Y}_i^s = \mathbf{y^*}$. First of all, we should recognize that if $A$ is not as strong as $A = \mathbf{y}_{-i}^s$, even the importance sampling techniques will not help much. See the discussion in the first paragraph of this section. Second of all, typically the set of inferred values of $\mathbf{Y}_i^s$, $\{\mathbf{y^*}\}$, is reduced to a much smaller set than the full set containing all possible combinations. Even though the reduction provides an upper bound on the attribute disclosure risks \citep{HuReiterWang2014PSD, HuReiterWang2018BA}, it is really for computational feasibility that such reduction is applied. 
Further research paths include faster algorithms to expand the small reduced set, and new algorithms to search for $\mathbf{y^*}$ that gives high probability estimation of $p(\mathbf{Z} \mid \mathbf{Y}_i^s = \mathbf{y^*}, \mathbf{y}^{us}, A, S)$ in an efficient way, therefore enabling the data disseminator to check against the actual truth $\mathbf{y}_i^s$ and determine its attribute disclosure risks level. Third of all, to use the Monte Carlo approximation coupled with importance sampling techniques in Equation (\ref{eq:AttriImportance3}), draws of $\Theta$ are necessary, which means the final synthetic data generation process involves parametric models. Among the selected examples, \citet{HuReiterWang2014PSD, WeiReiter2016} had parametric models for the outcome (multinomial and poisson, respectively). Even though \cite{WangReiter2012} used non-parametric CART synthesizers, their ultimate synthetic data generation process involves Bayesian bootstrap sampling with mixture normal distributions. It is unclear how to estimate the attribute disclosure risks for true non-parametric synthesizers, which can be a fruitful research path.

There are additional possible difficulties in implementing the Bayesian estimation procedure of attribute disclosure risks evaluation. As noted in \citet{ManriqueHu2018JRSSA}, their proposed synthesizers for categorical variables with structural zeros had serious stability issues with the estimation of $p(\mathbf{Z} \mid \mathbf{Y}_i^s = \mathbf{y^*}, \mathbf{y}^{us}, A, S)$, as its values varied by several thousands in the log-scale from one sample of $\Theta$ to another, resulting in enormous mean-squared error. The authors then developed an indirect bootstrap hypothesis testing framework to approximate the ranking of $\mathbf{y^*}$ in the reduced set. We refer the readers to \citet{ManriqueHu2018JRSSA} for details. 

One final comment to make is the work of \citet{McClureReiter2012DPandSyn}, where the authors compared the disclosure risk criterion of $\epsilon$-differential privacy with a criterion based on the attribute disclosure risk probabilities. The evaluation from their simulation studies was that the two paradigms are not easily reconciled. Moreover, sometimes attribute disclosure risks can be small even when $\epsilon$ is large. The authors proposed an alternative disclosure risk assessment approach, one integrates both paradigms, though great computation challenges were foreseeable. Further research on risk assessment integrating the two paradigms is desired.

\section{Bayesian estimation of identification disclosure risks}
\label{identificationrisk}

As discussed previously, we only consider identification disclosure risks for partially synthetic data.

Researchers have worked on Bayesian probabilistic matching to estimate the probabilities of identifications of sampled units. \citet{DuncanLambert1986, DuncanLambert1989, Lambert1993} developed Bayesian approaches to i) model the behavior of intruders, and ii) quantify sources of uncertainty about those estimated probabilities. Their work is followed by \citet{Fienberg1997continuous}, who estimated probabilities of identification for continuous microdata, which had undergone SDL techniques by adding random noise.

Observing the lack of illustrative applications on genuine data, \citet{Reiter2005risk} extended the Duncan-Lambert framework using data from the Current Population Survey (CPS). Common SDL techniques (recoding, topcoding, swapping, adding random noise, and combinations of these techniques) were applied to genuine microdata in their illustrations. They also considered different assumptions of intruders' knowledge and behavior and incorporated such information into the estimation of the identification probabilities. 

The step-by-step probability estimation procedure in \citet{Reiter2005risk} has been standard practice for Bayesian probabilistic matching ever since, especially after the synthetic data approach has gained its momentum. \citet{ReiterMitra2009} in particular first set up the framework for the Bayesian probabilistic matching for partially synthetic data. 

We now turn to the framework in \citet{ReiterMitra2009} for identification disclosure risks estimation for synthetic data, which was an extension of the general framework for identification disclosure risks estimation of data that have been subjected to common SDL techniques in \citet{Reiter2005risk}. We use similar notations, highlight the key steps, and illustrate with selected examples. We have chosen these examples that are built upon the framework but tailored for specific purposes and needs. To be as comprehensive as possible, we present two partially synthetic categorical data applications i) \citet{ReiterMitra2009}, ii) \citet{DrechslerHu2018}, and iii) partially synthetic categorical and continuous data \citep{DrechslerReiter2010}. In the end, we will discuss the challenges and future directions of this framework.

\subsection{Notations and setup}
\label{identificationrisk:setup}
In the sample $\textsl{S}$ of $n$ units and $p$ variables, the notation $y_{ij}$ refers to the $j$-th variable of the $i$-th unit, where $i = 1, \cdots, n$ and $j = 0, 1, \cdots, p$. The column $j=0$ contains some unique identifiers (such as name or Social Security Number), which are never released. Among the recorded variables, i) some are available to users from external databases, denoted by  $\mathbf{y}_i^A$, and ii) others are unavailable to users except in the released data, denoted by $\mathbf{y}_i^U$. We therefore have the vector response of the $i$-th unit, $\mathbf{y}_i = (y_{i1}, \cdots, y_{ip}) = (\mathbf{y}_i^A, \mathbf{y}_i^U)$. We also have the matrix $\mathbf{Y} = (\mathbf{Y}^{A}, \mathbf{Y}^{U})$ representing the original values of all $n$ units.

On the agency side, suppose it releases all $n$ units of the sample $\textsl{S}$. Similar to the split of $\mathbf{y}^i$, we have $\mathbf{z}_i = (z_{i1}, \cdots, z_{ip}) = (\mathbf{z}_i^A, \mathbf{z}_i^U)$. Among the available variables, we further split them into i) $\mathbf{z}_i^{A_s}$ the synthesized variables, and ii) $\mathbf{z}_i^{A_{us}}$ the un-synthesized variables. We therefore have $\mathbf{z}_i = (\mathbf{z}_i^{A_{us}}, \mathbf{z}_i^{A_s}, \mathbf{z}_i^U)$, and we let $\mathbf{Z} = (\mathbf{Z}^{A_{us}}, \mathbf{Z}^{A_s}, \mathbf{Z}^U)$ be the matrix of all released data. We also let $\mathbf{Y}^{A_s}$ be all $n$ unit's original values of the synthesized variables. We note that in some cases, the agency might only release $r \leq n$ units of the sample \citep{Reiter2005risk}.

On the intruder side, let $\mathbf{t}$ be the vector of information that the intruder has. $\mathbf{t}$ may or may not be in $\mathbf{Z}$, but we assume $\mathbf{t} = \mathbf{y}_i^A$ for some unit in the population. This vector $\mathbf{t}$ only contains un-synthesized and synthesized variables (no unavailable variables as in $\mathbf{y}_i$ and $\mathbf{z}_i$), thus we have $\mathbf{t} = (\mathbf{t}^{A_{us}}, \mathbf{t}^{A_s})$. The intruder's goal is to match record $i$ in $\mathbf{Z}$ to the target when $z_{i0} = t_0$. Additionally, two other pieces of information can be available to the intruder. Let $S$ represent the meta-data released about the simulation models used to generate the synthetic data, and let $R$ represent the meta-data released about the reason why records were selected for synthesis. Either $S$ or $R$ could be empty. 

There are $ n$ released units in $\mathbf{Z}$. Let $I$ be the random variable that equals to $i$ when $z_{i0} = t_0$ for $i \in \mathbf{Z}$ and equals $n + 1$ when $z_{i0} = t_0$ for some $i \notin \mathbf{Z}$. The intruder intends to calculate $Pr(I = i \mid \mathbf{t}, \mathbf{Z}, S, R)$ for $i = 1, \cdots, n+1$. The intruder is particularly interested in learning whether any of the calculated identification probabilities for $i = 1, \cdots, n$ are large enough to declare an identification.

For the agency, it is paramount to model different intruder's knowledge and behavior when estimating identification risks from releasing synthetic dataset. The framework in \citet{ReiterMitra2009} allows the incorporation of these different assumptions at multiple stages in the estimating process, thus gives extensive flexibility to parties trying to evaluate identification disclosure risks.

\subsection{Key estimating steps}
\label{identificationrisk:keysteps}

The intruder intends to calculate $Pr(I = i \mid \mathbf{t}, \mathbf{Z}, S, R) = $ for $i = 1, \cdots, n+1$. Based on the split of $\mathbf{Z} = (\mathbf{Z}^{A_{us}}, \mathbf{Z}^{A_s}, \mathbf{Z}^U)$, we re-write the probability as 

\begin{eqnarray}
Pr(I = i \mid \mathbf{t}, \mathbf{Z}^{A_{us}}, \mathbf{Z}^{A_s}, \mathbf{Z}^U, S, R). 
\label{eq:IndEq1}
\end{eqnarray}

In fact, the intruder does not know the actual values in $\mathbf{Y}^{A_s}$, all $n$ unit's original values of the synthesized variables. Therefore for the intruder, integrating over its possible values when computing the match probabilities is necessary, as in

\begin{eqnarray}
Pr(I = i \mid \mathbf{t}, \mathbf{Z}^{A_{us}}, \mathbf{Z}^{A_s}, \mathbf{Z}^U, S, R) &=& \int Pr(I = i \mid \mathbf{t}, \mathbf{Z}^{A_{us}}, \mathbf{Z}^{A_s}, \mathbf{Z}^U, S, R, \mathbf{Y}^{A_s}) \nonumber \\
&&Pr(\mathbf{Y}^{A_s} \mid \mathbf{t}, \mathbf{Z}^{A_{us}}, \mathbf{Z}^{A_s}, \mathbf{Z}^U, S, R) d \mathbf{Y}^{A_s}.
\label{eq:IndEq2}
\end{eqnarray}

The estimation procedure of Equation (\ref{eq:IndEq2}) varies by the variable(s) in $\mathbf{t}$ (e.g. whether in $\mathbf{t}^{A_{us}}$ or in $\mathbf{t}^{A_s}$), the variable types, assumptions on the level of knowledge of $\mathbf{t}$ being in $\mathbf{Z}$ or not, of $S$ and $R$, among other things. Here we go through each of these aspects/quantities and their implications in the estimating process, highlight several common practices that have been adopted, before we illustrate with a selection of identification disclosure risk assessment demonstrations with real synthetic data applications in Section \ref{identificationrisk:examples}.

\subsubsection{The variable(s) in $\mathbf{t}^{A_{us}}$}

An immediate simplification of $Pr(I = i \mid \mathbf{t}, \mathbf{Z}^{A_{us}}, \mathbf{Z}^{A_s}, \mathbf{Z}^U, S, R, \mathbf{Y}^{A_s})$ in Equation (\ref{eq:IndEq2}) is

\begin{eqnarray}
Pr(I = i \mid \mathbf{t}, \mathbf{Z}^{A_{us}}, \mathbf{Z}^{A_s}, \mathbf{Z}^U, S, R, \mathbf{Y}^{A_s}) = Pr(I = i \mid \mathbf{t}, \mathbf{Z}^{A_{us}}, \mathbf{Y}^{A_s}).
\label{eq:IndEq3}
\end{eqnarray}

This is true because when $\mathbf{Y}^{A_s}$ is given, $I = i$ and $\{\mathbf{Z}^{A_s}, \mathbf{Z}^U, S, R\}$ are conditionally independent. That is, the intruder would use $(\mathbf{Z}^{A_{us}}, \mathbf{Y}^{A_s})$ without the synthetic data $\mathbf{Z}^{A_s}$, the unavailable variables $\mathbf{Z}^{U}$, $S$, or $R$ to attempt re-identification. Equation (\ref{eq:IndEq3}) will be used in Sections \ref{identificationrisk:keysteps:2} and \ref{identificationrisk:keysteps:3} as well.

Consider any variable $k$ in $\mathbf{t}^{A_{us}}$. Since it is an un-synthesized variable, for any unit $i$ in $\mathbf{Z}^{A_{us}}$ where the released value of $z_{ik} \neq t_k$, $Pr(I = i \mid \mathbf{t}, \mathbf{Z}^{A_{us}}, \mathbf{Y}^{A_s}) = 0$.

\subsubsection{The variable(s) in $\mathbf{t}^{A_{s}}$}
\label{identificationrisk:keysteps:2}

For categorical variables in the synthesized set $\mathbf{t}^{A_s}$, the intruder matches directly on $\mathbf{Z}^{A_s}$. For numerical or continuous variables in $\mathbf{t}^{A_s}$, while exact match could be pursued, the nature of the numerical/continuous variables will result in zero probabilities for most if not all of the records. Therefore, it is advisable to match numerical components of $\mathbf{Z}^{A_s}$ within some acceptable distance (e.g. Euclidean or Mahalonobis) from the corresponding $\mathbf{t}^{A_s}$.

\subsubsection{Whether $\mathbf{t}$ is in $\mathbf{Z}$ or not}
\label{identificationrisk:keysteps:3}

The overall assumption we have is that the vector of information that the intruder has, $\mathbf{t} = \mathbf{y}_i^{A}$ for some unit in the population, but not necessarily in $\mathbf{Z}$. When $\mathbf{t} = \mathbf{y}_i^{A}$ is in $\mathbf{Z}$, then the quantity in Equation (\ref{eq:IndEq3}) for $I = n+1$ is 0, i.e. $Pr(I = n+1 \mid \mathbf{t}, \mathbf{Z}^{A_{us}}, \mathbf{Y}^{A_s}) = 0$. This simplifies calculating $Pr(I = i \mid \mathbf{t}, \mathbf{Z}^{A_{us}}, \mathbf{Y}^{A_s})$ for $i \leq n$. For example,
\begin{eqnarray}
Pr(I = i \mid \mathbf{t}, \mathbf{Z}^{A_{us}}, \mathbf{Y}^{A_s}) = \frac{1}{n_t},
\label{eq:IndEq4}
\end{eqnarray}
where $n_t$ is the number of units in $(\mathbf{Z}^{A_{us}}, \mathbf{Y}^{A_s})$ with $\mathbf{y}_i^{A}$ consistent with $\mathbf{t}$. 

When $\mathbf{t} = \mathbf{y}_i^{A}$ is not in $\mathbf{Z}$, then $Pr(I = n+1 \mid \mathbf{t}, \mathbf{Z}^{A_{us}}, \mathbf{Y}^{A_s}) \neq 0$. If we let $N_t$ be the number of units in the population that have $(\mathbf{z}_i^{A_{us}}, \mathbf{y}_i^{A_s})$ consistent with $\mathbf{t}$ which are also included in $\mathbf{Z}$, then 

\begin{eqnarray}
\begin{cases}
Pr(I = i \mid \mathbf{t}, \mathbf{Z}^{A_{us}}, \mathbf{Y}^{A_s}) = \frac{1}{N_t}, \,\,\, \\
Pr(I = n+1 \mid \mathbf{t}, \mathbf{Z}^{A_{us}}, \mathbf{Y}^{A_s}) = \frac{N_t - n_t}{N_t}. \\
\end{cases}
\label{eq:IndEq5}
\end{eqnarray}

Determining $N_t$ can be done from census totals, or can be estimated from available sources. \citet{ReiterMitra2009} discussed possible ways for estimation using survey weights. Model-based approaches to estimating $N_t$ can be applied too, for example \citet{ElamirSkinner2006JOS}, among others. Additional approaches to accounting for intruder uncertainty due to sampling were proposed in \citet{DrechslerReiter2008}.

It is important to recognize that setting $Pr(I = n+1 \mid \mathbf{t}, \mathbf{Z}^{A_{us}}, \mathbf{Y}^{A_s}) = 0$ results in conservative measures of identification disclosure risks.

\subsubsection{Assumptions about $S$ and $R$}
\label{identificationrisk:keysteps:4}
Previously, we let $S$ represent the meta-data released about the simulation models used to generate the synthetic data, and  $R$ represent the meta-data released about the reason why records were selected for synthesis. We note that in practice, $R$ is usually dropped because reasons why records were selected for synthesis are difficult to come by. However, $S$ can be available in many cases. For example, in Section \ref{attributerisk:keysteps}, information about the synthesis models of the SIPP Synthetic Beta is available online \citep{SIPPdocument}, which should be assumed to be known by the intruder. Not only the SIPP, information about the synthesis process of the SynLBD is publicly available in \citet{SynLBD2011, SynLBD2014}.

\subsubsection{Estimating $Pr(I = i \mid \mathbf{t}, \mathbf{Z}^{A_{us}}, \mathbf{Z}^{A_s}, \mathbf{Z}^U, S, R)$ through Monte Carlo}
\label{identificationrisk:keysteps:5}
This description follows the description given in \citet{DrechslerHu2018}. The construction in Equation (\ref{eq:IndEq2}) suggests a Monte Carlo approach to estimating each $Pr(I = i \mid \mathbf{t}, \mathbf{Z}, S$) (note that $\mathbf{Z}$ is used in place of $(\mathbf{Z}^{A_{us}}, \mathbf{Z}^{A_s}, \mathbf{Z}^{U})$; $R$ is dropped, assumed unavailable), and we re-write it as

\begin{eqnarray}
Pr(I = i \mid \mathbf{t}, \mathbf{Z}, S) = \int Pr(I = i \mid \mathbf{t}, \mathbf{Z}, S, \mathbf{Y}^{A_s}) Pr(\mathbf{Y}^{A_s} \mid \mathbf{t}, \mathbf{Z}, S) d \mathbf{Y}^{A_s}.
\label{eq:IndEq6}
\end{eqnarray}

For the Monte Carlo approach, perform the following two-step process.

\begin{enumerate}
\item[(i)] Sample a value of $\mathbf{Y}^{A_s}$ from $Pr(\mathbf{Y}^{A_s} \mid \mathbf{t}, \mathbf{Z}, S)$, and let $\mathbf{Y}_{new}$ represent one set of simulated values.

\item[(ii)] Compute $Pr(I = i \mid \mathbf{t}, \mathbf{Z}, S, \mathbf{Y}^{A_s} = \mathbf{Y}_{new})$ using exact matching assuming $\mathbf{Y}_{new}$ are collected values.
\end{enumerate}

This two-step process is iterated $h$ times, where ideally $h$ is large, and Equation (\ref{eq:IndEq6}) is estimated as

\begin{eqnarray}
Pr(I = i \mid \mathbf{t}, \mathbf{Z}, S) \approx \frac{1}{h}\sum_{h} Pr(I = i \mid \mathbf{t}, \mathbf{Z}, S, \mathbf{Y}^{A_s} = \mathbf{Y}_{new}^{(h)}),
\end{eqnarray}
where $\mathbf{Y}^{A_s} = \mathbf{Y}_{new}^{(h)}$ indicates one iteration of the two-step process.

When $S$ has no information, the intruder treats the simulated values as plausible draws of $\mathbf{Y}^{A_s}$.

\subsubsection{Three summaries of identification disclosure probabilities}
\label{identificationrisk:keysteps:6}
For attribute disclosure risk measures in Section \ref{attributerisk:examples}, summaries of attribute disclosure probabilities vary by variable types and contexts. For example, fully synthetic categorical data uses summaries of i) ranking, and ii) re-normalized probability of the true record being disclosed, as in \citet{HuReiterWang2014PSD} in Section \ref{attributerisk:examples:fullcat}. Partially synthetic continuous data, specifically in \citet{WangReiter2012} where synthetic precise geographies are released, summaries of i) the average Euclidean distance between the intruder's inferred value of the geography and the actual geography, and ii) the count of the actual cases in circle centered at the actual geography within radius in i) are reported. 

Unlike the summaries of attribute disclosure probabilities, summaries of identification disclosure probabilities are more generally applicable, regardless of the variable types and contexts. There are three summaries of identification disclosure probabilities, which now we describe, following \citet{DrechslerHu2018}.

We need the following notations and definitions before we present the three summaries. Let $c_i$ be the number of records with the highest match probability for the target $\mathbf{t}_i$; let $T_i = 1$ if the true match is among the $c_i$ units and $T_i = 0$ otherwise. Let $K_i = 1$ when $c_i T_i = 1$ and $K_i = 0$ otherwise, and let $N$ denote the total number of target records. Finally, let $F_i = 1$ when $c_i (1 - T_i) = 1$ and $F_i = 0$ otherwise, and let $s$ equal the number of records with $c_i = 1$.

Now we can present the three widely used summaries (file-level) of identification disclosure probabilities using the notations and definitions given above. 

\begin{enumerate}
\item[(i)] The expected match risk: 

\begin{eqnarray}
\sum_{i=1}^{n} \frac{T_i}{c_i}.
\end{eqnarray}

When $T_i = 1$ and $c_i > 1$, the contribution of unit $i$ to the expected match risk is the probability of finding the correct match by randomly selecting for the $c_i$ candidates. In general, the higher the expected match risk, the higher the identification disclosure risks.
\item[(ii)] The true match rate:

\begin{eqnarray}
\sum_{i=1}^{n} \frac{K_i}{N},
\end{eqnarray}
which is the percentage of true unique matches among the target records. In general, the higher the true match rate, the higher the identification disclosure risks.

\item[(iii)] The false match rate:

\begin{eqnarray}
\sum_{i=1}^{n} \frac{F_i}{s},
\end{eqnarray}
which is the percentage of false matches among unique matches. In general, the lower the false match rate, the higher the identification disclosure risks.

\end{enumerate}

\subsection{Selected examples}
\label{identificationrisk:examples}

In this selected examples section, we want to show the readers a few different applications of partially synthetic data. We will illustrate the variables in $\mathbf{t} = (\mathbf{t}^{A_{us}}, \mathbf{t}^{A_s})$ for each application. All applications follow similar estimating procedures, and report the same three summaries as presented in Section \ref{identificationrisk:keysteps:6}: i) the expected match risk, ii) the true match rate, and iii) the false match rate.

For each application, we give a brief overview of the dataset(s) and research questions to provide the background. We also mention the synthesizers, but the details of the synthesizers and the evaluation of the utility of the synthetic data are omitted. Interested readers should refer to the cited papers for further information.

\subsubsection{Partially synthetic categorical data 1}
\label{identificationrisk:examples:parcat1}

\citet{ReiterMitra2009} aimed at partially synthesizing a sample of $n = 2562$ of the 1987 Survey of Youth in Custody. There are 23 variables on the file, and the authors illustrated partially synthesizing two categorical variables, facility and race. Table \ref{tab:ReiterMitra2009} gives a partial list of the variables with their description, synthesis information and whether known by the intruder. All other un-listed 20 variables are not known by the intruder during the identification disclosure risks evaluation.
 
\begin{table}[bth]
\centering
\caption{Selected variables used in the \citet{ReiterMitra2009}. Data taken
  from the 1987 Survey of Youth in Custody. \label{tab:ReiterMitra2009}}
\begin{tabular}{llll}
\hline
Variable &  Description & Synthesized & Known by intruder\\ \hline
Facility & Categorical, 46 levels & Yes  & Yes\\
Race & Categorical, 5 levels & Yes & Yes\\
Ethnicity & Categorical, 2 levels & No & Yes \\ \hline
\end{tabular}
\end{table}

To synthesize the facility and race variables, the authors first use multinomial regressions to synthesize facility. All other variables except race and some variables causing multi-collinearity are included in the multinomial regressions as predictors. Once all values of the facility variable are synthesized, the authors then synthesize race using multinomial regressions. The predictors in these multinomial regressions include all other variables plus indicator variables for facilities, except those causing multi-collinearity. \citet{ReiterMitra2009} note that the new values of race are simulated conditional on the values of the synthetic facility indicators.

For the identification disclosure risks evaluation, the authors considered facility and race in $\mathbf{t}^{A_s}$, and ethnicity in $\mathbf{t}^{A_{us}}$. They also assumed that all targets are in the sample, i.e. $Pr(I = 2562 + 1 \mid \mathbf{t}, \mathbf{Z}^{A_{us}}, \mathbf{Y}^{A_s}) = 0$. 

\subsubsection{Partially synthetic categorical data 2}
\label{identificationrisk:examples:parcat2}

\citet{DrechslerHu2018} aimed at comparing a few existing synthesizers on a large German administrative database called the Integrated Employment Biographies (IEB) to provide access to detailed geocoding information. There are approximately 22 million records in the IEB. The authors considered 11 variables as listed in Table \ref{tab:DrechslerHu2018}. We include the variables, the description, whether a variable is synthesized, and whether the variable is known by the intruder in the identification disclosure risks estimation (i.e. whether in $\mathbf{t}$) in this table. The authors in fact experimented with different numbers of variables to be synthesized in order to provide higher protection. However, Table \ref{tab:DrechslerHu2018} considers the main synthesis approach, on which the authors presented most of the utility and risks results too.

\begin{table}[bth]
\centering
\caption{Variables used in the \citet{DrechslerHu2018}. Data taken
  from the IEB database in Germany. Note that the exact geocoding information is recorded as distance in meters from the point 52 northern latitude and 10 eastern longitude. It is converted to categorical for two out of the three synthesizers.\label{tab:DrechslerHu2018}}
\begin{tabular}{llll}
\hline
Variable &  Description & Synthesized & Known by intruder\\ \hline
Exact geocoding info & Longitude and latitude &Yes & Yes\\
Sex & Male, female & No  & Yes\\
Foreign & Yes, no & No  & Yes\\
Age & 6 categories & No & Yes \\
Education & 6 categories & No & No \\
Occupation level & 7 categories & No & No\\
Occupation & 12 categories & No & Yes \\
Industry of the employer & 15 categories & No & Yes\\
Wage & 10 categories (quantiles) & No & No\\
Distance to work & 5 categories & No & No \\
ZIP code & 2,063 ZIP code levels & No & No\\ \hline
\end{tabular}
\end{table}

The authors considered three synthesizers. The first synthesizer is the DPMPM synthesizer used in \citet{HuReiterWang2014PSD}, where the exact geocoding information was discretized into one unordered categorical variable, and the Dirichlet Process mixture model on the joint of the 11 unordered categorical variables was estimated and used to generate synthetic data. The second synthesizer is the CART synthesizer used in \citet{WangReiter2012}, where the exact geocoding information (the latitude and longitude) were treated as continuous and synthesized sequentially. We call this CART synthesizer the CART continuous. The third synthesizer is also a CART synthesizer, but similar to the DPMPM synthesizer, the exact geocoding information was discretized into one unordered categorical variable. We call this CART synthesizer the CART categorical. All three synthesizers were applied to generate partially synthetic IEB, where only the geocoding information was synthesized (either as categorical or as continuous). 

For the identification disclosure risks evaluation, the authors considered the exact geocoding information in $\mathbf{t}^{A_s}$, and sex, foreign, age, occupation, and industry of the employer in $\mathbf{t}^{A_{us}}$. Because the IEB is a census, the authors also assumed that all targets are in $\mathbf{Z}$, i.e. $Pr(I = n + 1 \mid \mathbf{t}, \mathbf{Z}^{A_{us}}, \mathbf{Y}^{A_s}) = 0$.  They reported the expected match risk, the true match rate, and the false match rate for different synthesizers. While the CART categorical synthesizer produced synthetic data with the highest utility, the identification disclosure risks may be deemed too high, therefore the authors recommended two approaches for increasing the level of protection: i) aggregate the geocoding information to a higher level, and ii) synthesize additional variables in the dataset. \citet{DrechslerHu2018} preferred ii) over i), and interested readers are referred to the paper for their discussion and general recommendations.

\subsubsection{Partially synthetic categorical and continuous data}
\label{identificationrisk:examples:parconcat}

\citet{DrechslerReiter2010} aimed at partially synthesizing a sample of $n = 51016$ of the March 2000 U.S. CPS. The authors in fact treated the sample as a census to illustrate their sampling with synthesis methodology, but for our illustration purpose, we will ignore the differences. There are 10 variables on the file, and the authors illustrated partially synthesizing three variables (2 are categorical and 1 is continuous). Table \ref{tab:DrechslerReiter2010} gives the list of the variables with their description, synthesis information and whether known by the intruder.

\begin{table}[bth]
\centering
\caption{Variables used in the \citet{DrechslerReiter2010}. Data taken
  from the March 2000 U.S. CPS. HH stands for household.\label{tab:DrechslerReiter2010}}
\begin{tabular}{llll}
\hline
Variable &  Description & Synthesized & Known by intruder\\ \hline
Sex & Male, female & No & Yes \\
Race & Categorical, 4 levels & Yes & Yes \\
Marital status & Categorical, 7 levels & Yes & Yes \\
Highest education level & Categorical, 16 levels & No & No \\
Age & Range 0-90 & Yes & Yes \\
\# of people in HH & 1-16 & No & No \\
\# of people in HH under 18 & 0-11 & No & No \\
Household property taxes & 0, 1-99,997 & No & No \\
Social security payments &  0, 1-50,000 & No & No \\
Household income & -21,011-768,742 & No & No \\ \hline
\end{tabular}
\end{table}

The authors synthesized age, race, and marital status sequentially using CART synthesizers. The synthetic values were generated through the Bayesian bootstrap.

For the identification disclosure risks evaluation, the authors considered race, martial status, and age in $\mathbf{t}^{A_{s}}$, and sex in $\mathbf{t}^{A_{us}}$. They reported the expected match risk, the true match rate, and the false match rate for a selected subset of 1089 sensitive cases.

\subsection{Discussion and comments}
\label{identificationrisk:discussion}
Current practices focus on the three summaries of identification disclosure probabilities (the expected match risk, the true match rate, and the false match rate). These are important and useful file-level measures. While important, there might be additional useful measures worth being developed. 

$R$, the meta-data released about the reason why records were selected for synthesis, is usually dropped because of lack of information. More simulation studies with different scenarios of $R$ could potentially provide insight on the effect of $R$ in identification disclosure risks assessment. 

The assumption about what the intruder knows (the ``Known by intruder" column in Tables \ref{tab:ReiterMitra2009}, \ref{tab:DrechslerHu2018}, \ref{tab:DrechslerReiter2010}) when performing matching can be arbitrary, and the calculated identification disclosure risks can be very sensitive to the assumption. To the best of our knowledge, no work on the sensitivity of identification disclosure risks to the intruder's knowledge assumption has been done, which is an important future work path.

Another important future work direction is matching with synthesized continuous attributes. As discussed in Section \ref{identificationrisk:keysteps:2}, matching numerical components of $\mathbf{Z}^{A_s}$ within some acceptable distance is preferred to exact matching. For example, if the continuous income attribute is synthesized, the intruder can declare a match if the synthesized income of a record is within some distance of an absolute value (e.g. 
\$10,000) from its original income value, or within some distance of a relative value (e.g. 20\% of the original income value) from its original income value. How to select the radius most likely depends on the statistical agency's threshold of what is considered as acceptable risks. Further research on the distance selection and evaluation of effects of the distance selection is needed to provide guidelines to data disseminators.

We primarily focus on estimating identification disclosure risks for synthetic data in this paper. However, the Duncan-Lambert framework developed in \citet{Reiter2005risk} can be applied to masked data that have been subjected to SDL techniques (e.g. data swapping, adding random noise, and micro-aggregation). Interested readers can refer to the general framework of \citet{Reiter2005risk} for further details of the estimation process and examples with SDL techniques.

\section{Summary}
\label{summary}

Estimating disclosure risks of synthetic data is an important aspect of evaluating the usefulness and feasibility of releasing publicly available synthetic microdata. In this paper, we present recent developments of Bayesian methods to disclosure risks assessment for both attribute disclosure and identification disclosure. We present the Bayesian thinking, highlight key steps, and illustrate with real synthetic data applications. We demonstrate the flexibility of Bayesian methods in modeling the beliefs of data intruders, as well as the interpretability of subjective Bayesian probabilities in terms of the disclosure risks evaluation. We also discuss and comment on limitations of current methods and some possible future research paths.

Our main discussion and comments have been presented separately for attribute disclosure and identification disclosure. However, there are some common points worth mentioning here. For one, we have exclusively presented on assessment processes based on the assumption that any record the intruder intends to make inference about is in the released sample. More research should be done to evaluate the uncertainty from sampling, especially for the fully synthetic case. For another, we have exclusively presented on assessment based on synthetic data. We would like to see more research on Bayesian methods for disclosure risks assessment based on other perturbation techniques, including the SDL techniques. We recognize that not all agencies are comfortable to generate and release synthetic microdata yet, therefore disclosure risks assessment based on other perturbation techniques is an important and promising research path.

Finally, there are other forms of risk that are not currently researched on extensively. For example, for both fully and partially synthetic data, one can ask the question ``can an intruder accurately infer from the synthetic data that some record with particular values is in the confidential data?" We believe such risk scenarios and other forms should be further researched on, providing a more comprehensive risk assessment of synthetic data, and more generally, of perturbed microdata.

\section*{Acknowledgements}
This research was supported by the NSF/ASA/BLS Research Fellowship Program and a Research Committee Award from Vassar College, NY, USA.

\bibliographystyle{natbib}
\bibliography{disclosurebib}

\end{document}